\documentclass[twocolumn,preprintnumbers]{revtex4}
\usepackage{amsmath,amssymb}
\usepackage{graphicx}
\usepackage{dcolumn}
\usepackage{epsfig,color}
\usepackage{bm}
\usepackage{verbatim}

\begin{document}

\title{Threshold Effects in Heavy Quarkonium Spectroscopy and Decays}

\author{J. Ferretti} 
\email[Corresponding author: ]{jacopo.ferretti@yale.edu}
\affiliation{Center for Theoretical Physics, Sloane Physics Laboratory, Yale University, New Haven, Connecticut 06520-8120, USA}

\begin{abstract}
The possible importance of threshold effects in heavy quarkonium spectroscopy is discussed. The starting point is the calculation of the spectrum of heavy quarkonium-like states with self-energy/threshold corrections. Two different approaches are compared: I) The Unquenched Quark Model (UQM); II) A novel coupled-channel model, based on the UQM formalism. The latter provides a possible solution to the long-standing problem of convergence in UQM calculations; it also makes it possible to distinguish between states which are almost pure quarkonia and exotic states, characterized by non-negligible threshold (or continuum) components in their wave functions. The UQM-based coupled-channel model is used to study the $\chi_{\rm c}(2P)$ and $\chi_{\rm b}(3P)$ multiplets: $\chi_{\rm c}(2P)$'s are described as charmonium-like states with non-negligible molecular-type components in their wave functions, $\chi_{\rm b}(3P)$'s as almost pure bottomonia. Other possible applications of the UQM and the UQM-based coupled-channel model formalisms to the calculation of other observables, like the strong decay amplitudes, are also discussed.
\end{abstract}

\maketitle

\section{Introduction}
\label{sec:intro}
In 2003 Belle discovered a meson, the $X(3872)$ [now $\chi_{\rm c1}(3872)$] \cite{Choi:2003ue,Acosta:2003zx,Abazov:2004kp}, whose properties are incompatible with a pure quark-antiquark interpretation.
The $X(3872)$ was the first example of an exotic hadron, namely a baryon/meson whose description requires the introduction of multiquark (tetra- and pentaquark) or gluonic degrees of freedom \cite{Tanabashi:2018oca,Esposito:2016noz,Olsen:2017bmm,Guo:2017jvc,Karliner:2017qhf,Liu:2019zoy}.
If we restrict to tetraquark or quarkonium-like exotics, several theoretical interpretations can be mentioned, including: a) Compact tetraquark (or diquark-antidiquark) states \cite{Jaffe:1976ih,Barbour:1979qi,Weinstein:1983gd,SilvestreBrac:1993ss,Brink:1998as,Maiani:2004vq,Barnea:2006sd,Santopinto:2006my,Ebert:2008wm,Deng:2014gqa,Zhao:2014qva,Anwar:2017toa,Hughes:2017xie,Wu:2016vtq,Agaev:2018vag,Liu:2019zuc,Wang:2019rdo}; b) Meson-meson molecular states \cite{Weinstein:1990gu,Manohar:1992nd,Tornqvist:1993ng,Martins:1994hd,Hanhart:2007yq,Thomas:2008ja,Baru:2011rs,Valderrama:2012jv,Aceti:2012cb,Guo:2013sya}; c) Structures resulting from kinematic or threshold effects caused by virtual particles \cite{Heikkila:1983wd,Pennington:2007xr,Li:2009ad,Danilkin:2010cc,Ortega:2012rs,Ferretti:2013faa,Ferretti:2013vua,Achasov:2015oia,Kang:2016jxw,Lu:2016mbb,Ferretti:2018tco}; Hadro-quarkonium states \cite{Dubynskiy:2008mq,Guo:2008zg,Wang:2009hi,Voloshin:2013dpa,Li:2013ssa,Wang:2013kra,Brambilla:2015rqa,Alberti:2016dru,Panteleeva:2018ijz,Ferretti:2018kzy}.
Here, we focus on the Unquenched Quark Model (UQM) \cite{Heikkila:1983wd,Pennington:2007xr,Li:2009ad,Danilkin:2010cc,Ortega:2012rs,Ferretti:2013faa,Ferretti:2013vua,Ferretti:2012zz,Ferretti:2014xqa,Bijker:2009up} description of heavy quarkonium-like states, which are interpreted as heavy quarkonium cores, $\left| A \right\rangle = \left| Q \bar Q \right\rangle$, plus meson-meson higher Fock (or continuum) components, $\left| BC \right\rangle = \left| Q \bar q - q \bar Q \right\rangle$, due to virtual particle/loop effects.

The UQM has a long story. Since its formulation in the early 80s \cite{Heikkila:1983wd}, a wide variety of applications has been explored, both in baryon/meson spectroscopy and structure. The UQM formalism is useful and elegant but, during the years, has shown some weaknesses and ambiguities. They include: I) The lack of convergence of its results when one considers large towers of meson-meson/meson-baryon intermediate states $\left| BC \right\rangle$; II) Ambiguities related to the nature of the $\left| BC \right\rangle$ continuum or higher Fock components, especially if the $\left| BC \right\rangle$ intermediate states are far from a meson-meson/meson-baryon threshold; III) Difficulties in the interpretation of the UQM results.

Hadron observables in the UQM are proportional to the ``propagator'' $\frac{1}{M_A - E_B - E_C}$, with $E_{B,C} =\sqrt{k^2 + M_{B,C}^2}$.
It can be easily shown that the form of the UQM pair-creation operator, which couples the $\left| A \right\rangle$ and $\left| BC \right\rangle$ states by creating a light quark-antiquark pair from the vacuum, cannot ensure a sharp distinction between intermediate states whose energies $E_B + E_C$ are close to $M_A$ or far from it.
Even the introduction of form factors in the pair-creation operator, in order to suppress $A$, $B$ and $C$ hadron wave function overlaps, do not provide a solution to the previous issue.
This strongly affects the convergence of the UQM.

To improve the reliability of the UQM, these difficulties should be urgently addressed. A possible solution to the previous inconsistencies is discussed in the present contribution. See also Refs. \cite{Ferretti:2018tco} and \cite{Ferretti:2019cyb}.
Other possible applications of the UQM and the UQM-based coupled-channel model formalisms to the calculation of other observables, like meson strong decay amplitudes, are discussed.

\section{Unquenching the quark model for heavy quarkonium-like mesons}
The procedure for ``unquenching the quark model'' relies on the introduction of higher Fock components in quark-antiquark bare meson wave functions. The first step is to introduce tetraquark or meson-meson molecular-type components, $\left| Q \bar q - q \bar Q \right\rangle$, where $Q = c$ or $b$ is a heavy quark and $q = u, d$ or $s$ a light one \cite{Heikkila:1983wd,Pennington:2007xr,Ortega:2012rs,Ferretti:2013faa,Ferretti:2013vua,Lu:2016mbb,Ferretti:2018tco,Ferretti:2012zz,Ferretti:2014xqa,Ferretti:2019cyb}.
The introduction of intermediate hybrid components, namely containing quark and gluonic degrees of freedom, will be discussed in the future.
The ``unquenched'' meson wave function can be written as
\begin{equation}	
	\label{eqn:Psi-A}
	\begin{array}{l}	
	\left| \psi_A \right\rangle = {\cal N} \left[ \left| A \right\rangle + \displaystyle \sum_{BC \ell J} \int k^2 dk
	\left| BC k \ell J \right\rangle \frac{ \left\langle BC k \ell J \right| T^{\dagger} \left| A \right\rangle}{M_A - E_B - E_C} \right] ~. 
	\end{array}
\end{equation}
Here, ${\cal N}$ is a normalization factor, $\left| \psi_A \right\rangle$ the superposition of a quark-antiquark configuration, $\left| A \right\rangle$, plus a sum over all the possible higher Fock components, $\left| BC \right\rangle$, due to the creation of a $^3P_0$ light $q \bar q$ pair. 
The sum is extended over a complete set of intermediate meson-meson states, $\left| BC \right\rangle$; $M_A$ is the physical mass of the meson $A$; $k$ and $\ell$ are the relative radial momentum and orbital angular momentum of $B$ and $C$, and $J$ is the total angular momentum, with ${\bf J} = {\bf J}_B + {\bf J}_C + {\bm \ell}$.
$T^{\dagger}$ in Eq. (\ref{eqn:Psi-A}) is the pair-creation operator of Refs. \cite{Ferretti:2013faa,Ferretti:2013vua,Ferretti:2018tco,Ferretti:2012zz,Ferretti:2014xqa,Ferretti:2019cyb}. See also Refs. \cite{Geiger:1991qe,Bijker:2009up}. 

Below threshold, $T^{\dagger}$ can couple a bare meson, $\left| A \right\rangle$, to four-quark meson-meson continuum components, $\left| BC \right\rangle$; above threshold, it can trigger $A \rightarrow BC$ open-flavor strong decays; see \cite{Ferretti:2015rsa} and references therein.
In the former case, the expectation value of a meson observable, $\hat {\mathcal O}_{\rm m}$, on $\left| \psi_A \right\rangle$, Eq. (\ref{eqn:Psi-A}), is given by
\begin{equation}
	\label{eqn:exp-value}
	\left\langle \psi_A \right| \hat {\mathcal O}_{\rm m} \left| \psi_A \right\rangle = \left\langle \hat {\mathcal O}_{\rm m} \right\rangle_{\rm val} 
	+ \left\langle \hat {\mathcal O}_{\rm m} \right\rangle_{\rm cont}  \mbox{ },
\end{equation}
where $\left\langle \hat {\mathcal O}_{\rm m} \right\rangle_{\rm val}$ and $\left\langle \hat {\mathcal O}_{\rm m} \right\rangle_{\rm cont}$ are the expectation values on the valence, $\left| A \right\rangle$, and continuum components, $\left| BC \right\rangle$, respectively.
Typical observables which can be calculated within the UQM formalism are the physical masses of quarkonium-like states, $M_A$, with self-energy corrections.
They are related to the bare and self-energies via
\begin{equation}
	\label{eqn:Ma-UQM}
	M_A = E_A + \Sigma(M_A)  \mbox{ },
\end{equation}
where the bare energies of pure quark-antiquark states, $E_A$, have to be evaluated in a specific quark model like, for example, the relativized QM of Ref. \cite{Godfrey:1985xj}.
In the UQM, the self-energies, $\Sigma(M_A)$, are computed according to
\begin{equation}
	\label{eqn:self-a}
	\Sigma(M_A) = \sum_{BC\ell J} \int_0^{\infty} k^2 dk \mbox{ } 
	\frac{\left| \left\langle BC k \ell J \right| T^\dag \left| A \right\rangle \right|^2}{M_A - E_B - E_C}  \mbox{ }.
\end{equation}
The UQM calculation of the heavy quarkonium spectrum via Eq. (\ref{eqn:Ma-UQM}) has problems of convergence.
Below, we discuss a simple procedure to ``renormalize'' the UQM results. More details can be found in Refs. \cite{Ferretti:2018tco} and \cite{Ferretti:2019cyb}.

\section{A novel coupled-channel approach based on the UQM formalism}
Here, we briefly describe the UQM-based coupled-channel approach of Ref. \cite{Ferretti:2018tco}.
In particular, we focus on the calculation of quarkonium-like meson masses with threshold corrections. This formalism can be easily used to compute other observables without major modifications.

The main difference with respect to the standard UQM approach of Refs. \cite{Ferretti:2013faa,Ferretti:2013vua} is that the coupled-channel model of Ref. \cite{Ferretti:2018tco} is not used to fit the whole heavy quarkonium spectrum, but it is applied to a single quarkonium multiplet at a time. Some examples include the $\chi_{\rm c}(2P)$ and $\chi_{\rm b}(3P)$ multiplets, which were studied in Ref. \cite{Ferretti:2018tco}.
Moreover, a ``subtraction'' or ``renormalization'' prescription for the UQM results is introduced \cite{Ferretti:2018tco}.
This consists in the substitution of Eq. (\ref{eqn:Ma-UQM}) with
\begin{equation}
	\label{eqn:new-Ma}
	M_A = E_A + \Sigma(M_A) + \Delta \mbox{ },
\end{equation}
where $\Delta$ is the only one free parameter for each multiplet.
It is defined as the smallest self-energy correction (in terms of absolute value) of a multiplet member \cite{Ferretti:2018tco}.
Its importance and impact on the UQM results are clarified by the examples provided below.

\subsection{Threshold mass-shifts in $\chi_{\rm c}(2P)$ and $\chi_{\rm b}(3P)$ multiplets}
As a possible application of the UQM-based coupled-channel model, we briefly discuss the results of Ref. \cite{Ferretti:2018tco} for the masses of $\chi_{\rm c}(2P)$ and $\chi_{\rm b}(3P)$ heavy quarkonia with threshold corrections.

\begin{figure}
\includegraphics[width=7cm]{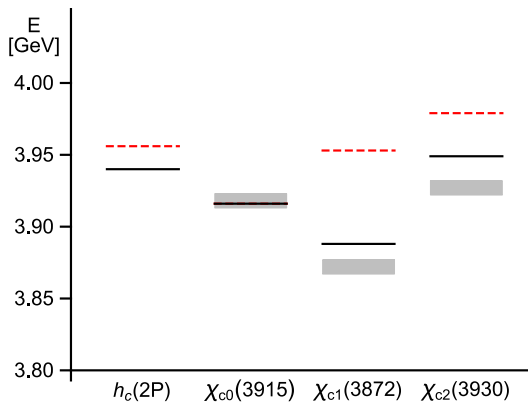}
\caption{$\chi_{\rm c}(2P)$ multiplet: masses with threshold corrections. Boxes, dashed and continuous lines correspond to the experimental \cite{Tanabashi:2018oca}, calculated bare and physical masses, respectively. See Table \ref{tab:ChiC(2P)-splittings} and Ref. \cite{Ferretti:2018tco}.}
\label{fig:ChiC(2P)-splittings}
\end{figure}

The threshold mass shifts were computed by considering a complete set of $1S 1S$ meson-meson intermediate states, like $D \bar D$, $D \bar D^*$ ($B \bar B$, $B \bar B^*$), and so on.
If one is interested in the study of a different quarkonium multiplet, like the $\chi_{\rm c}(3P)$, one should substitute the previous $1S 1S$ loops with the set of meson-meson intermediate states which is closer to the multiplet of interest, like $1S1P$ or $1S2S$.
The values of the bare masses, $E_A$, were directly extracted from the relativized model of Refs. \cite{Godfrey:1985xj,Barnes:2005pb,Godfrey:2015dia}, without refitting the potential model parameters to the spectrum via Eq. (\ref{eqn:Ma-UQM}); the values of the physical meson masses, $M_A$, were taken from the PDG \cite{Tanabashi:2018oca}.
The UQM parameters, used in the calculation of the $\left\langle BC k \ell J \right| T^\dag \left| A \right\rangle$ vertices of Eq. (\ref{eqn:self-a}), were extracted from Refs. \cite{Ferretti:2015rsa,Ferretti:2013faa,Ferretti:2013vua}. 
Finally, our UQM-based coupled-channel model results for the self-energy/threshold corrections and physical masses of $\chi_{\rm c}(2P)$ and $\chi_{\rm b}(3P)$ multiplets are reported in Table \ref{tab:ChiC(2P)-splittings} and Figure \ref{fig:ChiC(2P)-splittings}.
%%%%%%%%%%%%%%%%%%%%%%%%%%%%%%%%%%%%%%%%%%%%%%%%%%
\begin{table*}
\caption{Comparison between the experimental masses \cite{Tanabashi:2018oca} of $\chi_{\rm c}(2P)$ and $\chi_{\rm b}(3P)$ states, $M_A^{\rm exp}$, and theoretical predictions from Ref. \cite{Ferretti:2018tco}, $M_A^{\rm th}$. The bare masses, $E_A$, are taken from Refs. \cite{Godfrey:1985xj,Barnes:2005pb,Godfrey:2015dia}. The experimental results denoted by $\dag$ are from Ref. \cite{Godfrey:2015dia}, where $\chi_{\rm b}(3P)$ predicted multiplet mass splittings were used in combination with the experimental value of the $\chi_{\rm b1}(3P)$ mass. Due to the lack of experimental data, in the $h_{\rm c}(2P)$ case we use the same value for the physical mass as the bare one \cite{Godfrey:1985xj}.}
\label{tab:ChiC(2P)-splittings}
\begin{ruledtabular}
\begin{tabular}{lcccc}
State  & $E_A$ [MeV]  & $\Sigma(M_A) + \Delta$ [MeV] & $M_A^{\rm th}$ [MeV] & $M_A^{\rm exp}$ [MeV] \\
\hline
$h_{\rm c}(2P)$       & 3956               & $-16$                                      & 3940                            & -- \\
$\chi_{\rm c0}(3915)$ or $\chi_{\rm c0}(2P)$ & 3916               & 0                                             & 3916                            & 3918  \\
$\chi_{\rm c1}(3872)$ or $\chi_{\rm c1}(2P)$ & 3953               & $-65$                                      & 3888                            & 3872  \\
$\chi_{\rm c2}(3930)$ or $\chi_{\rm c2}(2P)$ & 3979               & $-30$                                      & 3949                            & 3927  \\ 
\hline
$h_{\rm b}(3P)$       & 10541            & $-4$                                         & 10538                          & 10519$^\dag$ \\
$\chi_{\rm b0}(3P)$ & 10522             & 0                                              & 10522                          & 10500$^\dag$  \\
$\chi_{\rm b1}(3P)$ & 10538             & $-2$                                         & 10537                          & 10513  \\
$\chi_{\rm b2}(3P)$ & 10550             & $-7$                                         & 10543                          & 10524  \\ 
\end{tabular}
\end{ruledtabular}
\end{table*}
%%%%%%%%%%%%%%%%%%%%%%%%%%%%%%%%%%%%%%%%%%%%%%%%%%

It is worth to observe that: I) Our theoretical predictions agree with the data within the error of a QM calculation, of the order of $30-50$ MeV; II) Among $\chi_{\rm c}(2P)$ multiplet members, the $\chi_{\rm c1}(2P)$ is subject to the largest threshold correction, $-65$ MeV, which brings its bare mass value, 3953 MeV, towards the experimental one, 3871.69 MeV \cite{Tanabashi:2018oca}; III) In the $\chi_{\rm c}(2P)$ case, threshold effects break the peculiar mass pattern of a $\chi$-type multiplet, i.e. $M_{\chi_0} < M_{\chi_1} \approx M_{\rm h} < M_{\chi_2}$. On the contrary, the previous mass pattern is respected in the $\chi_{\rm b}(3P)$ case; IV) Threshold effects are negligible in the $\chi_{\rm b}(3P)$ case. Because of this, we interpret $\chi_{\rm b}(3P)$ states as almost pure bottomonia.  
This purely bottomonium description of $\chi_{\rm b}(3P)$ states agrees with the experimental masses of the $\chi_{\rm b1}(3P)$ and $\chi_{\rm b2}(3P)$; the latter was recently reported by the PDG \cite{Tanabashi:2018oca}, $10524.02\pm0.57\pm0.53$ MeV.
%%%%%%%%%%%%%%%%%%%%%%%%%%%%%%%%%%%%%%%%%%%%%%%%%%
%%%%%%%%%%%%%%%%%%%%%%%%%%%%%%%%%%%%%%%%%%%%%%%%%%

\subsection{Continuum components}
\label{Continuum components}
The norm of the continuum (or molecular-type) component of a quarkonium-like state is given by \cite{Ferretti:2018tco,Ferretti:2012zz,Ferretti:2014xqa}
\begin{equation}
	\label{eqn:Pa-sea}
	P_A^{\rm cont} = \sum_{BC\ell J} \int_0^\infty q^2 dq \mbox{ } 
	\frac{\left| \left\langle BC q  \, \ell J \right| T^\dag \left| A \right\rangle \right|^2}{(M_A - E_B - E_C)^2}  \mbox{ },
\end{equation}
where the probability to find the meson in its valence component, $P_A^{\rm val}$, is given by $P_A^{\rm val} = 1 - P_A^{\rm cont}$.
In the coupled-channel model of Ref. \cite{Ferretti:2018tco}, the procedure to compute the norm of Eq. (\ref{eqn:Pa-sea}) is slightly different than in the UQM \cite{Ferretti:2012zz,Ferretti:2014xqa}, as discussed in the following.

For each multiplet member, we define the ratio
\begin{equation}
	\label{eqn:Ri}
	R_i = \frac{\Sigma(M_i) - \Delta}{\Sigma(M_i)}  \mbox{ },
\end{equation}
where $\Sigma(M_i)$ and $\Delta$ are defined in Eqs. (\ref{eqn:self-a}) and (\ref{eqn:new-Ma}), and the index $i = 1, ..., N_{\rm mult}$ runs over the meson multiplet members. For each member of the multiplet, we also define an effective and renormalized pair-creation strength
\begin{equation}
	\label{eqn:gamma0i-ren}
	\tilde \gamma_{0,i}^{\rm eff} = \gamma_0^{\rm eff} \sqrt{R_i}  \mbox{ },
\end{equation}
where the effective pair-creation strength, $\gamma_0^{\rm eff}$, is that of \cite{Ferretti:2012zz}, Eq. (12).
We can now calculate the norm of the continuum (or molecular-type) component for each multiplet member by means of Eq. (\ref{eqn:Pa-sea}) and replacing the effective pair-creation strength, $\gamma_0^{\rm eff}$, with that of Eq. (\ref{eqn:gamma0i-ren}).

As an example, we give results for the normalization of the continuum and valence components of the $X(3872)$ in Table \ref{tab:continuum}. See also Ref. \cite{Ferretti:2018tco}.
%%%%%%%%%%%%%%%%%%%%%%%%%%%%%%%%%%%%%%%%%%%%%%%%%%
\begin{table}
\caption{Calculated probabilities to find the $X(3872)$ in its valence, $P_A^{\rm val}$, or meson-meson continuum component, $P_A^{\rm cont}$; see Eq. (\ref{eqn:Pa-sea}) and Ref. \cite{Ferretti:2018tco}. The two probabilities are related via $P_A^{\rm val} = 1 - P_A^{\rm cont}$.}
\label{tab:continuum}
\begin{ruledtabular}
\begin{tabular}{ccc}
State                                                     & Component        & Probability \\
\hline
$X(3872)$ or $\chi_{\rm c1}(3872)$      & Continuum                          & 0.853  \\    
                                                              & Valence                                      & 0.147  \\ 
\end{tabular}
\end{ruledtabular}
\end{table}
%%%%%%%%%%%%%%%%%%%%%%%%%%%%%%%%%%%%%%%%%%%%%%%%%%

\section{Heavy quarkonium strong decays in the UQM-based coupled channel model formalism}
\begin{figure}
\includegraphics[width=9cm]{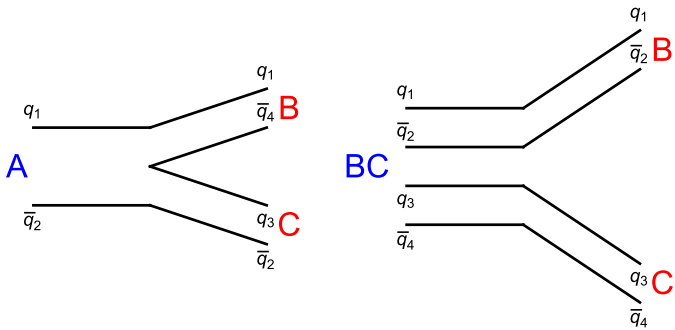}
\caption{Diagrams which contribute to the $A \rightarrow BC$ open-flavor strong decay width at the leading order. On the left, the decay of the $q_1 \bar q_2$ valence component takes place via the creation of a light quark-antiquark pair, $q_3 \bar q_4$. On the right, the decay of the $q_1 \bar q_2 q_3 \bar q_4$ molecular type component proceeds by dissociation of the $BC$ meson-meson molecular-type state.}
\label{fig:UQM-decays}
\end{figure}

The decay amplitudes of heavy quarkonia can be calculated in the UQM \cite{Ferretti:2013faa,Ferretti:2013vua} or the UQM-based coupled channel model formalism of Ref. \cite{Ferretti:2018tco} by making use of Eq. (\ref{eqn:exp-value}):
\begin{equation}
	\Gamma_A^{\rm UQM} = P_A^{\rm val} \Gamma_A^{\rm val} 
	+ \sum_{BC} P_{BC}^{\rm cont} \Gamma_{BC}^{\rm cont}  \mbox{ }.
\end{equation}
In the particular case of open-flavor strong decays, one has
\begin{equation}
	\Gamma_A^{\rm UQM} = P_A^{\rm val} \Gamma_{A \rightarrow BC}^{\rm val} 
	+ \sum_{BC} P_{BC}^{\rm cont} \Gamma_{BC \rightarrow BC}^{\rm cont}  \mbox{ },
\end{equation}	
where $P_A^{\rm val}$ and $P_{BC}^{\rm cont}$ are the probability to find the meson $A$ in its valence and $BC$ continuum or molecular-type components, respectively. $\Gamma_{A \rightarrow BC}^{\rm val}$ is the $A \rightarrow BC$ open-flavor strong decay width of Figure \ref{fig:UQM-decays}, left panel, which can be computed in the $^3P_0$ pair-creation model \cite{Micu:1968mk,LeYaouanc:1972vsx,Kokoski:1985is,Capstick:1993kb,Barnes:2005pb,Ferretti:2015rsa}; $\Gamma_{BC \rightarrow BC}^{\rm cont}$ is the $BC \rightarrow BC$ dissociation width of the molecular-type component $BC$, whose diagram is depicted in Figure \ref{fig:UQM-decays}, right panel.
The calculation of the open-flavor strong decay amplitudes of heavy quarkonia in the UQM-based coupled channel model formalism of Ref. \cite{Ferretti:2018tco} will be the subject of a subsequent paper.
%%%%%%%%%%%%%%%%%%%%%%%%%%%%%%%%%%%%%%%%%%%%%%
%%%%%%%%%%%%%%%%%%%%%%%%%%%%%%%%%%%%%%%%%%%%%%

Below, we briefly discuss a calculation of the $J/\psi \omega$ and $J/\psi \rho$ hidden-flavor strong decays of the $X(3872)$ in the UQM \cite{Ferretti:2018tco}. In this particular case, the decay width has no contribution associated with the valence component.

\subsection{Hidden-flavor strong decays of the $X(3872)$ in the UQM}
In the UQM, the $J/\psi \omega$ and $J/\psi \rho$ hidden-flavor transitions of the $X(3872)$ can be seen as two-step processes \cite{Ferretti:2018tco}.
Firstly, the initially pure $c \bar c$ state, $\chi_{c1}(2^3P_1)$, is ``dressed'' with $1S1S$ open-charm molecular-type components, including $D \bar D$, $D \bar D^*$, ..., by means of the UQM formalism \cite{Ferretti:2013faa,Ferretti:2013vua,Ferretti:2012zz,Ferretti:2014xqa}. 
The wave function of the $X(3872)$ is thus made up of a $\chi_{c1}(2^3P_1)$ core, $\left| A \right\rangle$, plus open-charm $D \bar D$, $D \bar D^*$, ... higher Fock components, $\left| BC \right\rangle$. 
Secondly, the $D^0 \bar D^{0*}$ molecular-type component of the $X(3872)$ dissociates into $J/\psi$ plus $\rho$ or $\omega$, which are indicated as $\left| D \right\rangle$ and $\left| E \right\rangle$ in Eq. (\ref{eqn:loop}).
The $\sigma_{BC \rightarrow DE}$ dissociation cross-sections are calculated by means of the non-relativistic formalism of Refs. \cite{Barnes-Swanson}.

By analogy with the positronium case \cite{Itzykson:1980rh}, the hidden-flavor decay width is given by \cite{Ferretti:2018tco}
\begin{widetext}
\begin{equation}
\label{eqn:loop}
	\begin{array}{l}
		\Gamma = \left[\displaystyle \int P_B^2dP_B \mbox{ } \frac{\left|\left\langle A| T^\dag |BC \right\rangle\right|^2}
		{(M_A - E_{BC})^2 + \frac{\Gamma_A^2}{4}}  \left| {\bf v}_B - {\bf v}_C \right| 
		\sigma_{BC \rightarrow DE}  \right]  \left|\Psi_{BC}(0)\right|^2    
	\end{array}	\mbox{ }.
\end{equation}
\end{widetext}
The term in square brackets in Eq. (\ref{eqn:loop}) is the product of the $BC \rightarrow DE$ dissociation cross-section, $\sigma_{BC \rightarrow DE}$, and $\left| {\bf v}_B - {\bf v}_C \right|$, which is the difference between the velocities of the mesons $B$ and $C$. The previous quantity is convoluted with a distribution function, which describes the probability to find the $\left| BC\right\rangle = \left| D^0 \bar D^{0*} \right\rangle$ component in the wave function of the $X(3872)$.
The term in square brackets is multiplied by $\left|\Psi_{BC}(0)\right|^2$, which is the squared wave function of the $BC$ molecular-type component, evaluated in the origin.

Finally, the results of the UQM calculation of Ref. \cite{Ferretti:2018tco} are: $\Gamma^{\rm UQM}_{X(3872) \rightarrow J/\psi \rho} = 10 \mbox{ keV }$, $\Gamma^{\rm UQM}_{X(3872) \rightarrow J/\psi \omega} = 6 \mbox{ keV }$.
It is worth noting that the ratio between the previous calculated amplitudes, 
\begin{equation}
	\label{eqn:ratio-th}
	\frac{\Gamma^{\rm UQM}_{X(3872) \rightarrow J/\psi \omega}}{\Gamma^{\rm UQM}_{X(3872) \rightarrow J/\psi \rho}} 
	= 0.6 \mbox{ },
\end{equation}
is compatible with the present experimental data \cite{Tanabashi:2018oca,delAmoSanchez:2010jr},
\begin{equation}
	\label{eqn:ratio-exp}
	\frac{\Gamma^{\rm exp}_{X(3872) \rightarrow J/\psi \omega}}{\Gamma^{\rm exp}_{X(3872) \rightarrow J/\psi \rho}} 
	= 0.8\pm0.3  \mbox{ },
\end{equation}
within the experimental error.

\section{Conclusion}
In this contribution, we discussed the possible importance of threshold effects in heavy quarkonium spectroscopy. We compared two different approaches: I) The Unquenched Quark Model (UQM) \cite{Ferretti:2013faa,Ferretti:2013vua}; II) A novel coupled-channel model, based on the UQM formalism \cite{Ferretti:2018tco}. The latter suggests a possible solution to the problem of convergence in UQM calculations; it also allows to distinguish between states which are almost pure quarkonia and exotic states, whose wave functions contain non-negligible threshold (or continuum) components. The coupled channel approach was used to study the quark structure of the $\chi_{\rm c}(2P)$ and $\chi_{\rm b}(3P)$ multiplets: the former was described as a multiplet of charmonium-like states with non-negligible molecular-type components in their wave functions, the latter as an almost pure bottomonium multiplet \cite{Ferretti:2018tco}. 

Other possible applications of the model, including the calculation of the strong decay amplitudes of quarkonium-like states, were also discussed. As an example, we summarized the UQM calculation of the $J/\psi \omega$ and $J/\psi \rho$ hidden-flavor strong decays of the $X(3872)$ of Ref. \cite{Ferretti:2018tco}.

\begin{acknowledgments}
The author acknowledges financial support from the US Department of Energy Grant No. DE-FG-02-91ER-40608.
\end{acknowledgments}

%\nocite{*}


\begin{thebibliography}{}
%\bibliography{aipsamp}% Produces the bibliography via BibTeX.

\bibitem{Choi:2003ue}
  S.~K.~Choi {\it et al.}  [Belle Collaboration],
  %``Observation of a narrow charmonium - like state in exclusive B+- ---> K+- pi+ pi- J / psi decays,''
  Phys.\ Rev.\ Lett.\  {\bf 91}, 262001 (2003).
  
\bibitem{Acosta:2003zx}
  D.~Acosta {\it et al.}  [CDF Collaboration],
  %``Observation of the narrow state $X(3872) \to J/\psi \pi^+ \pi^-$ in $\bar{p}p$ collisions at $\sqrt{s} = 1.96$ TeV,''
  Phys.\ Rev.\ Lett.\  {\bf 93}, 072001 (2004).  
  
\bibitem{Abazov:2004kp}
  V.~M.~Abazov {\it et al.}  [D0 Collaboration],
  %``Observation and properties of the $X(3872)$ decaying to $J/\psi \pi^+ \pi^-$ in $p\bar{p}$ collisions at $\sqrt{s} = 1.96$ TeV,''
  Phys.\ Rev.\ Lett.\  {\bf 93}, 162002 (2004).    
  
\bibitem{Tanabashi:2018oca} 
  M.~Tanabashi {\it et al.} [Particle Data Group],
  %``Review of Particle Physics,''
  Phys.\ Rev.\ D {\bf 98}, 030001 (2018).    
  
\bibitem{Esposito:2016noz}
  A.~Esposito, A.~Pilloni and A.~D.~Polosa,
  %``Multiquark Resonances,''
  Phys.\ Rept.\  {\bf 668}, 1 (2016).

\bibitem{Olsen:2017bmm}
  S.~L.~Olsen, T.~Skwarnicki and D.~Zieminska,
  %``Nonstandard heavy mesons and baryons: Experimental evidence,''
  Rev.\ Mod.\ Phys.\  {\bf 90}, 015003 (2018).  
  
\bibitem{Karliner:2017qhf} 
  M.~Karliner, J.~L.~Rosner and T.~Skwarnicki,
  %``Multiquark States,''
  Ann.\ Rev.\ Nucl.\ Part.\ Sci.\  {\bf 68}, 17 (2018).  
  
\bibitem{Guo:2017jvc} 
  F.~K.~Guo, C.~Hanhart, U.~G.~Mei{\ss}ner, Q.~Wang, Q.~Zhao and B.~S.~Zou,
  %``Hadronic molecules,''
  Rev.\ Mod.\ Phys.\  {\bf 90}, 015004 (2018).  
  
\bibitem{Liu:2019zoy} 
  Y.~R.~Liu, H.~X.~Chen, W.~Chen, X.~Liu and S.~L.~Zhu,
  %``Pentaquark and Tetraquark states,''
  Prog. Part. Nucl. Phys. {\bf 107}, 237 (2019). 
  
\bibitem{Jaffe:1976ih}
  R.~L.~Jaffe,
  %``Multi-Quark Hadrons. 2. Methods,''
  Phys.\ Rev.\ D {\bf 15}, 281 (1977).

\bibitem{Barbour:1979qi}
  I.~M.~Barbour and D.~K.~Ponting,
  %``Nonstrange Four Quark States,''
  Z.\ Phys.\ C {\bf 5}, 221 (1980);
  I.~M.~Barbour and J.~P.~Gilchrist,
  %``The $N \bar{N}$ and $\pi \pi$ Decay Modes of Baryonium,''
  Z.\ Phys.\ C {\bf 7}, 225 (1981)
  Erratum: [Z.\ Phys.\ C {\bf 8}, 282 (1981)].

\bibitem{Weinstein:1983gd}
  J.~D.~Weinstein and N.~Isgur,
  %``The q q anti-q anti-q System in a Potential Model,''
  Phys.\ Rev.\ D {\bf 27}, 588 (1983).	

\bibitem{SilvestreBrac:1993ss}
  B.~Silvestre-Brac and C.~Semay,
  %``Systematics of L = 0 q-2 anti-q-2 systems,''
  Z.\ Phys.\ C {\bf 57}, 273 (1993).

\bibitem{Brink:1998as}
  D.~M.~Brink and F.~Stancu,
  %``Tetraquarks with heavy flavors,''
  Phys.\ Rev.\ D {\bf 57}, 6778 (1998).  
  
\bibitem{Maiani:2004vq}
  L.~Maiani, F.~Piccinini, A.~D.~Polosa and V.~Riquer,
  %``Diquark-antidiquarks with hidden or open charm and the nature of X(3872),''
  Phys.\ Rev.\ D {\bf 71}, 014028 (2005).   
  
\bibitem{Barnea:2006sd}
  N.~Barnea, J.~Vijande and A.~Valcarce,
  %``Four-quark spectroscopy within the hyperspherical formalism,''
  Phys.\ Rev.\ D {\bf 73}, 054004 (2006).

\bibitem{Santopinto:2006my}
  E.~Santopinto and G.~Galat\`a,
  %``Spectroscopy of tetraquark states,''
  Phys.\ Rev.\ C {\bf 75}, 045206 (2007).
  
\bibitem{Ebert:2008wm} 
  D.~Ebert, R.~N.~Faustov, V.~O.~Galkin and W.~Lucha,
  %``Masses of tetraquarks with two heavy quarks in the relativistic quark model,''
  Phys.\ Rev.\ D {\bf 76}, 114015 (2007);
  D.~Ebert, R.~N.~Faustov and V.~O.~Galkin,
  %``Relativistic description of heavy tetraquarks,''
  Phys.\ Atom.\ Nucl.\  {\bf 72}, 184 (2009).  
  
\bibitem{Deng:2014gqa} 
  C.~Deng, J.~Ping and F.~Wang,
  %``Interpreting $Z_c(3900)$ and $Z_c(4025)/Z_c(4020)$ as charged tetraquark states,''
  Phys.\ Rev.\ D {\bf 90}, 054009 (2014).  
  
\bibitem{Zhao:2014qva} 
  L.~Zhao, W.~Z.~Deng and S.~L.~Zhu,
  %``Hidden-Charm Tetraquarks and Charged $Z_c$ States,''
  Phys.\ Rev.\ D {\bf 90}, 094031 (2014).  
  
\bibitem{Anwar:2017toa} 
  M.~N.~Anwar, J.~Ferretti, F.~K.~Guo, E.~Santopinto and B.~S.~Zou,
  %``Spectroscopy and decays of the fully-heavy tetraquarks,''
  Eur.\ Phys.\ J.\ C {\bf 78}, 647 (2018);
  M.~N.~Anwar, J.~Ferretti and E.~Santopinto,
  %``Spectroscopy of the hidden-charm $[qc][\bar q \bar c]$ and $[sc][\bar s \bar c]$ tetraquarks in the relativized diquark model,''
  Phys.\ Rev.\ D {\bf 98}, 094015 (2018).  
  
\bibitem{Hughes:2017xie} 
  C.~Hughes, E.~Eichten and C.~T.~H.~Davies,
  %``Searching for beauty-fully bound tetraquarks using lattice nonrelativistic QCD,''
  Phys.\ Rev.\ D {\bf 97}, 054505 (2018).  
  
\bibitem{Wu:2016vtq} 
  J.~Wu, Y.~R.~Liu, K.~Chen, X.~Liu and S.~L.~Zhu,
  %``Heavy-flavored tetraquark states with the $QQ\bar{Q}\bar{Q}$ configuration,''
  Phys.\ Rev.\ D {\bf 97}, 094015 (2018). 
  
\bibitem{Agaev:2018vag} 
  S.~S.~Agaev, K.~Azizi, B.~Barsbay and H.~Sundu,
  %``The doubly charmed pseudoscalar tetraquarks $T_{cc;\bar{s} \bar{s}}^{++}$ and $T_{cc;\bar{d} \bar{s}}^{++}$,''
  Nucl.\ Phys.\ B {\bf 939}, 130 (2019).   
  
\bibitem{Liu:2019zuc} 
  M.~S.~Liu, Q.~F.~L\"u, X.~H.~Zhong and Q.~Zhao,
  %``All-heavy tetraquarks,''
  Phys.\ Rev.\ D {\bf 100}, 016006 (2019).  
  
\bibitem{Wang:2019rdo} 
  G.~J.~Wang, L.~Meng and S.~L.~Zhu,
  %``Spectrum of the fully-heavy tetraquark state $QQ\bar Q' \bar Q'$,''
  arXiv:1907.05177.  
  
\bibitem{Weinstein:1990gu}
  J.~D.~Weinstein and N.~Isgur,
  %``K anti-K Molecules,''
  Phys.\ Rev.\ D {\bf 41}, 2236 (1990).

\bibitem{Manohar:1992nd}
  A.~V.~Manohar and M.~B.~Wise,
  %``Exotic Q Q anti-q anti-q states in QCD,''
  Nucl.\ Phys.\ B {\bf 399}, 17 (1993).

\bibitem{Tornqvist:1993ng}
  N.~A.~T\"ornqvist,
  %``From the deuteron to deusons, an analysis of deuteron - like meson meson bound states,''
  Z.\ Phys.\ C {\bf 61}, 525 (1994);
  %``Isospin breaking of the narrow charmonium state of Belle at 3872-MeV as a deuson,''
  Phys.\ Lett.\ B {\bf 590}, 209 (2004).
  
\bibitem{Martins:1994hd} 
  K.~Martins, D.~Blaschke and E.~Quack,
  %``Quark exchange model for charmonium dissociation in hot hadronic matter,''
  Phys.\ Rev.\ C {\bf 51}, 2723 (1995);
  C.~Pe\~na and D.~Blaschke,
  %``$X(3872)$ as a $D-\bar{D}^*$ molecule bound by quark exchange forces,''
  Acta Phys.\ Polon.\ Supp.\  {\bf 5}, 963 (2012).  

\bibitem{Hanhart:2007yq}
  C.~Hanhart, Y.~S.~Kalashnikova, A.~E.~Kudryavtsev and A.~V.~Nefediev,
  %``Reconciling the X(3872) with the near-threshold enhancement in the D0 anti-D*0 final state,''
  Phys.\ Rev.\ D {\bf 76}, 034007 (2007).

\bibitem{Thomas:2008ja}
  C.~E.~Thomas and F.~E.~Close,
  %``Is X(3872) a molecule?,''
  Phys.\ Rev.\ D {\bf 78}, 034007 (2008).

\bibitem{Baru:2011rs}
  V.~Baru, A.~A.~Filin, C.~Hanhart, Y.~S.~Kalashnikova, A.~E.~Kudryavtsev and A.~V.~Nefediev,
  %``Three-body $D\bar{D}\pi$ dynamics for the X(3872),''
  Phys.\ Rev.\ D {\bf 84}, 074029 (2011).

\bibitem{Valderrama:2012jv}
  M.~P.~Valderrama,
  %``Power Counting and Perturbative One Pion Exchange in Heavy Meson Molecules,''
  Phys.\ Rev.\ D {\bf 85}, 114037 (2012).
  
\bibitem{Aceti:2012cb}
  F.~Aceti, R.~Molina and E.~Oset,
  %``The $X(3872) \to J/\psi \gamma$ decay in the $D \bar D^*$ molecular picture,''
  Phys.\ Rev.\ D {\bf 86}, 113007 (2012).  

\bibitem{Guo:2013sya}
  F.~K.~Guo, C.~Hidalgo-Duque, J.~Nieves and M.~P.~Valderrama,
  %``Consequences of Heavy Quark Symmetries for Hadronic Molecules,''
  Phys.\ Rev.\ D {\bf 88}, 054007 (2013).  
  
\bibitem{Heikkila:1983wd}
  K.~Heikkila, S.~Ono and N.~A.~Tornqvist,
  %``HEAVY c anti-c AND b anti-b QUARKONIUM STATES AND UNITARITY EFFECTS,''
  Phys.\ Rev.\ D {\bf 29}, 110 (1984)
  Erratum: [Phys.\ Rev.\ D {\bf 29}, 2136 (1984)].  
  
\bibitem{Pennington:2007xr}
  M.~R.~Pennington and D.~J.~Wilson,
  %``Decay channels and charmonium mass-shifts,''
  Phys.\ Rev.\ D {\bf 76}, 077502 (2007).  
  
\bibitem{Li:2009ad}
  B.~-Q.~Li, C.~Meng and K. T.~Chao,
  %``Coupled-Channel and Screening Effects in Charmonium Spectrum,''
  Phys.\ Rev.\ D {\bf 80}, 014012 (2009).    
  
\bibitem{Danilkin:2010cc}
  I.~V.~Danilkin and Y.~A.~Simonov,
  %``Dynamical origin and the pole structure of X(3872),''
  Phys.\ Rev.\ Lett.\  {\bf 105}, 102002 (2010).    
  
\bibitem{Ortega:2012rs}
  P.~G.~Ortega, J.~Segovia, D.~R.~Entem and F.~Fernandez,
  %``Coupled channel approach to the structure of the X(3872),''
  Phys.\ Rev.\ D {\bf 81}, 054023 (2010); 
  P.~G.~Ortega, D.~R.~Entem and F.~Fernandez,
  %``Molecular Structures in Charmonium Spectrum: The $XYZ$ Puzzle,''
  J.\ Phys.\ G {\bf 40}, 065107 (2013).  
  
\bibitem{Ferretti:2013faa} 
  J.~Ferretti, G.~Galat\`a and E.~Santopinto,
  %``Interpretation of the X(3872) as a charmonium state plus an extra component due to the coupling to the meson-meson continuum,''
  Phys.\ Rev.\ C {\bf 88}, 015207 (2013).   		      
  
\bibitem{Ferretti:2013vua}   
  J.~Ferretti and E.~Santopinto,
  %``Higher mass bottomonia,''
  Phys.\ Rev.\ D {\bf 90}, 094022 (2014).    
  
\bibitem{Achasov:2015oia} 
  N.~N.~Achasov and E.~V.~Rogozina,
  %``X(3872), I^G(J^{PC})=0^+(1^{++}), as the \chi_{1c}(2P) charmonium,''
  Mod.\ Phys.\ Lett.\ A {\bf 30}, 1550181 (2015).    
  
\bibitem{Kang:2016jxw} 
  X.~W.~Kang and J.~A.~Oller,
  %``Different pole structures in line shapes of the $X(3872)$,''
  Eur.\ Phys.\ J.\ C {\bf 77}, 399 (2017).  
  
\bibitem{Lu:2016mbb} 
  Y.~Lu, M.~N.~Anwar and B.~S.~Zou,
  %``Coupled-Channel Effects for the Bottomonium with Realistic Wave Functions,''
  Phys.\ Rev.\ D {\bf 94}, 034021 (2016);
  M.~N.~Anwar, Y.~Lu and B.~S.~Zou,
  %``$\chi_{b}(3P)$ Multiplet Revisited: Ultrafine Mass Splitting and Radiative Transitions,''
  Phys.\ Rev.\ D {\bf 99}, 094005 (2019).      
  
\bibitem{Ferretti:2018tco} 
  J.~Ferretti and E.~Santopinto,
  %``Threshold corrections of $\chi_{\rm c}(2P)$ and $\chi_{\rm b}(3P)$ states in a coupled-channel model,''
  Phys.\ Lett.\ B {\bf 789}, 550 (2019).    
  
\bibitem{Dubynskiy:2008mq} 
  S.~Dubynskiy and M.~B.~Voloshin,
  %``Hadro-Charmonium,''
  Phys.\ Lett.\ B {\bf 666}, 344 (2008).  
  
\bibitem{Guo:2008zg} 
  F.~K.~Guo, C.~Hanhart and U.~G.~Mei{\ss}ner,
  %``Evidence that the Y(4660) is a f(0)(980)psi-prime bound state,''
  Phys.\ Lett.\ B {\bf 665}, 26 (2008); 
  %``Implications of heavy quark spin symmetry on heavy meson hadronic molecules,''
  Phys.\ Rev.\ Lett.\  {\bf 102}, 242004 (2009).  
  
\bibitem{Wang:2009hi} 
  Z.~G.~Wang and X.~H.~Zhang,
  %``Analysis of Y(4660) and related bound states with QCD sum rules,''
  Commun.\ Theor.\ Phys.\  {\bf 54}, 323 (2010);
  %``Analysis of the pseudoscalar partner of the Y(4660) and related bound states,''
  Eur.\ Phys.\ J.\ C {\bf 66}, 419 (2010).  
  
\bibitem{Voloshin:2013dpa} 
  M.~B.~Voloshin,
  %``$Z_c(3900)$ - what is inside?,''
  Phys.\ Rev.\ D {\bf 87}, 091501 (2013).
  
\bibitem{Li:2013ssa} 
  X.~Li and M.~B.~Voloshin,
  %``$Y$(4260) and $Y$(4360) as mixed hadrocharmonium,''
  Mod.\ Phys.\ Lett.\ A {\bf 29}, 1450060 (2014).  
  
\bibitem{Wang:2013kra} 
  Q.~Wang, M.~Cleven, F.~K.~Guo, C.~Hanhart, U.~G.~Mei{\ss}ner, X.~G.~Wu and Q.~Zhao,
  %``Y(4260): hadronic molecule versus hadro-charmonium interpretation,''
  Phys.\ Rev.\ D {\bf 89}, 034001 (2014);
  M.~Cleven, F.~K.~Guo, C.~Hanhart, Q.~Wang and Q.~Zhao,
  %``Employing spin symmetry to disentangle different models for the XYZ states,''
  Phys.\ Rev.\ D {\bf 92}, 014005 (2015).    
  
\bibitem{Brambilla:2015rqa} 
  N.~Brambilla, G.~Krein, J.~Tarr\'us Castell\`a and A.~Vairo,
  %``Long-range properties of $1S$ bottomonium states,''
  Phys.\ Rev.\ D {\bf 93}, 054002 (2016).    
  
\bibitem{Alberti:2016dru} 
  M.~Alberti, G.~S.~Bali, S.~Collins, F.~Knechtli, G.~Moir and W.~S\"oldner,
  %``Hadroquarkonium from lattice QCD,''
  Phys.\ Rev.\ D {\bf 95}, 074501 (2017).  
  
\bibitem{Panteleeva:2018ijz} 
  J.~Y.~Panteleeva, I.~A.~Perevalova, M.~V.~Polyakov and P.~Schweitzer,
  %``Tetraquarks with hidden charm and strangeness as $\phi$-$\psi$(2S) hadrocharmonium,''
  Phys.\ Rev.\ C {\bf 99}, 045206 (2019).      
  
\bibitem{Ferretti:2018kzy} 
  J.~Ferretti,
  %``$\eta_c$- and $J/\psi$-isoscalar meson bound states in the hadro-charmonium picture,''
  Phys.\ Lett.\ B {\bf 782}, 702 (2018);
  %``Effective Degrees of Freedom in Baryon and Meson Spectroscopy,''
  Few Body Syst.\  {\bf 60}, 17 (2019);
  J.~Ferretti, E.~Santopinto, M.~Naeem Anwar and M.~A.~Bedolla,
  %``The baryo-quarkonium picture for hidden-charm and bottom pentaquarks and LHCb $P_{\rm c}(4380)$ and $P_{\rm c}(4450)$ states,''
  Phys.\ Lett.\ B {\bf 789}, 562 (2019).
  
\bibitem{Bijker:2009up} 
  E.~Santopinto and R.~Bijker,
  %``Quark-antiquark effects in baryons,''
  Few Body Syst.\  {\bf 44}, 95 (2008);
  %``Flavor asymmetry of sea quarks in the unquenched quark model,''
  Phys.\ Rev.\ C {\bf 82}, 062202 (2010);
  R.~Bijker and E.~Santopinto,
  %``Unquenched quark model for baryons: Magnetic moments, spins and orbital angular momenta,''
  Phys.\ Rev.\ C {\bf 80}, 065210 (2009);
  E.~Santopinto, R.~Bijker and J.~Ferretti,
  %``Unquenching the quark model,''
  Few Body Syst.\  {\bf 50}, 199 (2011);
  R.~Bijker, J.~Ferretti and E.~Santopinto,
  %``$s\bar{s}$ sea pair contribution to electromagnetic observables of the proton in the unquenched quark model,''
  Phys.\ Rev.\ C {\bf 85}, 035204 (2012);
  E.~Santopinto, H.~Garc\'ia-Tecocoatzi and R.~Bijker,
  %``Electroproduction of baryon?meson states and strangeness suppression,''
  Phys.\ Lett.\ B {\bf 759}, 214 (2016)
  H.~Garc\'ia-Tecocoatzi, R.~Bijker, J.~Ferretti and E.~Santopinto,
  %``Self-energies of octet and decuplet baryons due to the coupling to the baryon-meson continuum,''
  Eur.\ Phys.\ J.\ A {\bf 53}, 115 (2017).
  
\bibitem{Ferretti:2012zz} 
  J.~Ferretti, G.~Galat\`a, E.~Santopinto and A.~Vassallo,
  %``Bottomonium self-energies due to the coupling to the meson-meson continuum,''
  Phys.\ Rev.\ C {\bf 86}, 015204 (2012).  
  
\bibitem{Ferretti:2014xqa} 
  J.~Ferretti, G.~Galat\`a and E.~Santopinto,
  %``Quark structure of the $X(3872)$ and $\chi_b(3P)$ resonances,''
  Phys.\ Rev.\ D {\bf 90}, 054010 (2014).  
  
\bibitem{Ferretti:2019cyb} 
  J.~Ferretti,
  %``Threshold effects in heavy quarkonium spectroscopy,''
  arXiv:1902.02835.    
  
\bibitem{Geiger:1991qe}
  P.~Geiger and N.~Isgur,
  %``The Quenched Approximation in the Quark Model,''
  Phys.\ Rev.\ D {\bf 41}, 1595 (1990); 
  %``How the Okubo-Zweig-Iizuka rule evades large loop corrections,''
  Phys.\ Rev.\ Lett.\  {\bf 67}, 1066 (1991).  
  
\bibitem{Ferretti:2015rsa} 
  R.~Bijker, J.~Ferretti, G.~Galat\`a, H.~Garc\'ia-Tecocoatzi and E.~Santopinto,
  %``Strong decays of baryons and missing resonances,''
  Phys.\ Rev.\ D {\bf 94}, 074040 (2016);
  H.~Garc\'ia-Tecocoatzi, R.~Bijker, J.~Ferretti, G.~Galat\`a and E.~Santopinto,
  %``Open flavor strong decays,''
  Few Body Syst.\  {\bf 57}, 985 (2016);
  J.~Ferretti and E.~Santopinto,
  %``Open-flavor strong decays of open-charm and open-bottom mesons in the $^3P_0$ model,''
  Phys.\ Rev.\ D {\bf 97}, 114020 (2018);
  E.~Santopinto, A.~Giachino, J.~Ferretti, H.~Garc\'ia-Tecocoatzi, M.~A.~Bedolla, R.~Bijker and E.~Ortiz-Pacheco,
  %``The $\Omega_{c}$-puzzle solved by means of spectrum and strong decay amplitude predictions,''
  arXiv:1811.01799.
  
\bibitem{Godfrey:1985xj}
  S.~Godfrey and N.~Isgur,
  %``Mesons in a Relativized Quark Model with Chromodynamics,''
  Phys.\ Rev.\ D {\bf 32}, 189 (1985).  
  
\bibitem{Barnes:2005pb} 
  T.~Barnes, S.~Godfrey and E.~S.~Swanson,
  %``Higher charmonia,''
  Phys.\ Rev.\ D {\bf 72}, 054026 (2005).  
    
\bibitem{Godfrey:2015dia} 
  S.~Godfrey and K.~Moats,
  %``Bottomonium Mesons and Strategies for their Observation,''
  Phys.\ Rev.\ D {\bf 92}, 054034 (2015).  
  
\bibitem{Micu:1968mk} 
  L.~Micu,
  %``Decay rates of meson resonances in a quark model,''
  Nucl.\ Phys.\ B {\bf 10}, 521 (1969).
  
\bibitem{LeYaouanc:1972vsx} 
  A.~Le Yaouanc, L.~Oliver, O.~Pene and J.~C.~Raynal,
  %``Naive quark pair creation model of strong interaction vertices,''
  Phys.\ Rev.\ D {\bf 8}, 2223 (1973).   
  
\bibitem{Kokoski:1985is} 
  R.~Kokoski and N.~Isgur,
  %``Meson Decays by Flux Tube Breaking,''
  Phys.\ Rev.\ D {\bf 35}, 907 (1987).    
  
\bibitem{Capstick:1993kb} 
  S.~Capstick and W.~Roberts,
  %``N pi decays of baryons in a relativized model,''
  Phys.\ Rev.\ D {\bf 47}, 1994 (1993);
  %``Quasi two-body decays of nonstrange baryons,''
  {\bf 49}, 4570 (1994).    
  
\bibitem{Barnes-Swanson}
  T.~Barnes and E.~S.~Swanson,
  Phys.\ Rev.\ D {\bf 46}, 131 (1992);
  T.~Barnes, E.~S.~Swanson, C.~Y.~Wong and X.~M.~Xu,
  Phys.\ Rev.\ C {\bf 68}, 014903 (2003).   
  
\bibitem{Itzykson:1980rh} 
  C.~Itzykson and J.~B.~Zuber,
  {\it Quantum Field Theory}, McGraw-Hill (1980).   
  
\bibitem{delAmoSanchez:2010jr} 
  P.~del Amo Sanchez {\it et al.}  [BaBar Collaboration],
  %``Evidence for the decay X(3872) ---> J/ psi omega,''
  Phys.\ Rev.\ D {\bf 82}, 011101 (2010).       
  
\end{thebibliography}
\end{document}